\documentclass[prl,aps,twocolumn,superscriptaddress]{revtex4-1}
\usepackage[T1]{fontenc}
\usepackage[utf8]{inputenc}

\usepackage{graphicx}
\usepackage{amssymb,amsfonts,amsmath}
\usepackage{color}
\usepackage{ulem}
\usepackage{bbm}
\usepackage{bbold}
\usepackage[hidelinks]{hyperref}
\usepackage{pdfpages}
\makeatletter
\AtBeginDocument{\let\LS@rot\@undefined}
\makeatother
\usepackage{pgffor}

\usepackage{tikz}
\usetikzlibrary{shapes}
\usetikzlibrary{patterns}
\usetikzlibrary{angles, quotes}

\newcommand{\rv}{{\mathbf r}}

\newcommand{\msphantom}[1]{$\ldots$}

\newcommand{\eqr}[1]{Eq.~\eqref{#1}}

\newcommand{\mydelete}[1]{{}}

\newcommand{\rmexc}{{\rm exc}}
\newcommand{\rmext}{{\rm ext}}

\newcommand{\rmid}{{\rm id}}

\newcommand{\rmr}{{\rm ref}}
\begin{document}

\title{Neural density functionals: Local learning and pair-correlation matching}

\author{Florian Samm\"uller}
\affiliation{Theoretische Physik II, Physikalisches Institut, 
  Universit{\"a}t Bayreuth, D-95447 Bayreuth, Germany}
\email{Florian.Sammueller@uni-bayreuth.de}

\author{Matthias Schmidt}
\affiliation{Theoretische Physik II, Physikalisches Institut, 
  Universit{\"a}t Bayreuth, D-95447 Bayreuth, Germany}
\email{Matthias.Schmidt@uni-bayreuth.de}

\date{5 June 2024, revised version: 25 July 2024}

\begin{abstract} Recently Dijkman {\it et
    al.}~(\href{https://doi.org/10.48550/arXiv.2403.15007}{arXiv:2403.15007})
  proposed training classical neural density functionals via bulk
  pair-correlation matching. We show their method to be an efficient regularizer
  for neural functionals based on local learning of inhomogeneous one-body
  direct correlations [Samm\"uller {\it et
    al.},~\href{https://doi.org/10.1073/pnas.2312484120} {Proc.\ Natl.\ Acad.\ Sci.\
    {\bf 120}, e2312484120 (2023)}]. While Dijkman {\it et al.}~demonstrated
  pair-correlation matching of a global neural free energy
  functional, we argue in favor of local one-body learning for flexible neural
  modelling of the full Mermin-Evans density functional map. Using spatial
  localization gives access to accurate neural free energy functionals,
  including convolutional neural networks, that transcend the training box.
\end{abstract}

\maketitle

As machine learning and density functional theory \cite{evans1979,
  hansen2013, evans1992, evans2016} share high computational
efficiency, it is natural to use neural networks to construct workable
and accurate approximations for the required density functional
relationships in the classical \cite{lin2019ml, lin2020ml, cats2022ml,
  yatsyshin2022, malpica-morales2023, dijkman2024ml,
  delasheras2023perspective, sammueller2023neural,
  sammueller2023whyNeural, sammueller2024hyperDFT, zimmermann2024ml,
  simon2023mlPatchy} and quantum realms \cite{nagai2018, jschmidt2018,
  zhou2019, nagai2020, li2021prl, li2022natcompsci, pederson2022,
  gedeon2022}. This common goal to make progress is nevertheless
approached from quite different directions and a considerable range of
different machine learning strategies have been put forward
\cite{lin2019ml, lin2020ml, cats2022ml, yatsyshin2022,
  malpica-morales2023, dijkman2024ml, delasheras2023perspective,
  sammueller2023neural, sammueller2023whyNeural,
  sammueller2024hyperDFT, zimmermann2024ml, simon2023mlPatchy,
  nagai2018, jschmidt2018, zhou2019, nagai2020, li2021prl,
  li2022natcompsci, pederson2022, gedeon2022, kelley2024}. The task is both
important and challenging. Overcoming the limited availability of
flexible classical density functional approximations would open up the
study of a much broader class of soft matter systems than is currently
accessible via, e.g., the hard sphere perturbation paradigm of
fundamental-measure theory \cite{rosenfeld1989, roth2006WhiteBear,
  roth2010} combined with mean-field attraction, as used in recent
work~\cite{evans2019pnas, coe2022prl, coe2023}.

In the classical context, typically the excess free energy functional
$F_\rmexc[\rho]$ is the object chosen to be approximated
\cite{evans1979, hansen2013}. We recall that $F_\rmexc[\rho]$ is the
nontrivial contribution to the total grand potential functional
$\Omega[\rho]=F_\rmid[\rho]+F_\rmexc[\rho]+\int d\rv
\rho(\rv)[V_\rmext(\rv)-\mu]$, where the ideal gas free energy
functional $F_\rmid[\rho]$ is known explicitly, $V_\rmext(\rv)$ is the
external potential that generates spatial inhomogeneity and $\mu$ is
the chemical potential that together with absolute temperature $T$
determines the thermodynamic conditions.  The Mermin-Evans variational
principle \cite{evans1979,mermin1965} ascertains that $\Omega[\rho]$
is minimized by the true equilibrium density profile $\rho(\rv)$, as a
function of position~$\rv$ across the system, and that the minimum
gives the equilibrium value of the grand potential.

The effects of the interparticle interactions are contained in
$F_\rmexc[\rho]$, which per se is not easily accessible and requires
the development of careful modelling strategies. Neural networks suit
this task very well, due to the clearcut correspondence between the
spatially resolved input density profile~$\rho(\rv)$ and the excess
free energy $F_\rmexc[\rho]$ as a global output value.  Reference data
for supervised machine learning is provided by many-body simulations
\cite{frenkel2023book, wilding2001, brukhno2021dlmonte}, which hence
offers the perspective of making theoretical predictions with
simulation precision.

Two very recent implementations use neural networks to represent the
functional relationship of either the global excess free energy
$F_\rmexc[\rho]$ \cite{dijkman2024ml} or the closely related one-body
direct correlation functional $c_1(\rv;[\rho])$
\cite{sammueller2023neural, sammueller2023whyNeural}.  The training
strategy in Ref.~\cite{dijkman2024ml} employs bulk pair-correlation
matching, while Refs.~\cite{sammueller2023neural,
  sammueller2023whyNeural} utilize a local learning approach involving
inhomogeneous one-body profiles. An appeal of the
former~\cite{dijkman2024ml} is the sole requirement of simulation
input in the form of bulk radial distribution functions $g(r)$ at
different densities for the particular model under
investigation. Local learning \cite{sammueller2023whyNeural,
  sammueller2023neural} rather requires inhomogeneous density profiles
as training data. Going beyond mere interpolation tasks, the resulting
neural functionals~\cite{sammueller2023whyNeural,
  sammueller2023neural} were however shown to be fit for carrying out deep
functional calculus based on exact sum rules and on functional
identities. Both methods were argued to be numerically highly accurate
\cite{dijkman2024ml, sammueller2023whyNeural, sammueller2023neural}.

In this Letter we contrast and cross-fertilize the underlying concepts
of pair-correlation matching \cite{dijkman2024ml} and of inhomogeneous
one-body learning \cite{sammueller2023whyNeural,
  sammueller2023neural}. We show how to incorporate pair-correlation
matching in a local learning strategy, which then retains the ability
of transcending beyond the box size of the training simulations.  In
keeping with statistical mechanics, pair-correlation matching
constitutes a physics-based regularizer for the training loss, thereby
avoiding potential problems of more generic regularization methods
\cite{chollet2017}.

We also demonstrate based on comparison to
reference simulation data and on violation of internal sum rule
consistency, that training density-functional dependence solely with
bulk densities and pair-correlation matching can yield unreliable neural performance in general
inhomogeneous situations.
We rationalize this behaviour in terms of the structure of the underlying exact functional dependencies.
We show that local learning does not suffer from such deficiencies and that the concept also
generalizes beyond constructing neural one-body direct correlations,
laying out two explicit strategies that make $F_\rmexc[\rho]$ directly
accessible.

We first describe key ideas of pair-correlation matching as proposed
by Dijkman {\it et al.}~\cite{dijkman2024ml} to construct a neural
network that represents the functional $F_\rmexc[\rho]$. Their method
exploits that $F_\rmexc[\rho]$ is a generating functional for the
hierarchy of direct correlation functionals. Specifically, at second
order the two-body direct correlation functional is obtained
\cite{hansen2013, evans1979}:
\begin{align}
  c_2(\rv,\rv';[\rho])  
  &= -\frac{\delta^2 \beta F_\rmexc[\rho]}
  {\delta\rho(\rv)\delta\rho(\rv')},
  \label{EQc2fromFexc}
\end{align}
where $\delta/\delta\rho(\rv)$ indicates the functional derivative
with respect to $\rho(\rv)$ and $\beta=1/(k_BT)$ with Boltzmann
constant $k_B$.  When specializing to the important, yet restricted,
case of bulk fluids, the density profile is constant,
$\rho(\rv)=\rho_b=\rm const$, and the bulk pair direct correlation
function $c_2(r;\rho_b)$ of liquid state theory is
recovered~\cite{hansen2013}. This constitutes a functional reduction
\begin{align}
 c_2(r; \rho_b) &= c_2(\rv,\rv';[\rho=\rho_b]),
 \label{EQc2reduction}
\end{align}
where the spatial dependence on $\rv$ and $\rv'$ simplifies to the
dependence on radial distance $r=|\rv-\rv'|$ on the left hand
side. Also the general functional dependence on the entirety of the
density profile $\rho(\rv)$ in \eqr{EQc2fromFexc} is reduced to a mere
parametric dependence on the value of the bulk density $\rho_b$ in
\eqr{EQc2reduction}.

Pair-correlation matching \cite{dijkman2024ml} exploits that results
for pair correlation (or radial distribution) functions $g(r)$ are
readily accessible via simulations. Using the total correlation
function $h(r)=g(r)-1$, the bulk Ornstein-Zernike equation ascertains
that $h(r) = c_2^\rmr(r) + \rho_b c_2^\rmr(r) \star h(r)$, where the
star indicates spatial convolution. We have denoted the bulk two-body
direct correlation function from this route by $c_2^\rmr(r)$, as
originating from the simulation data for $g(r)$. Numerical solution of
the Ornstein-Zernike equation yields quasi-exact reference results
for~$c_2^\rmr(r)$; details are given in the
SI~\cite{sammueller2024pairmatchingSI}. Pair-correlation matching
\cite{dijkman2024ml} trains $F_\rmexc[\rho]$ as a convolutional neural
network by minimizing the difference of $c_2(r;\rho_b)$, as obtained
via \eqr{EQc2fromFexc} and functional reduction~\eqref{EQc2reduction},
against the quasi-exact reference data for $c_2^\rmr(r)$ from
simulations and Ornstein-Zernike inversion.  Thereby the functional
derivative in \eqr{EQc2fromFexc} is performed via automatic
differentiation (autodiff) \cite{dijkman2024ml, sammueller2023neural,
  sammueller2023whyNeural, sammueller2024hyperDFT,
  baydin2018autodiff}. The aim is to achieve equality,
\begin{align}
  c_2(r;\rho_b) = c_2^\rmr(r; \rho_b).
  \label{EQc2matching}
\end{align}
Crucially, during training the neural functional is only evaluated via
its Hessian with respect to the input in \eqr{EQc2reduction} at
constant density profiles $\rho(\rv)=\rho_b$ with varying values of
$\rho_b$. No inhomogeneous density profiles are encountered during
training.

By contrast local direct correlation learning
\cite{sammueller2023neural, sammueller2023whyNeural} starts with the
standard liquid state relationship
\begin{align}
  c_1^\rmr(\rv) = \ln\rho(\rv) + \beta V_\rmext(\rv) - \beta \mu,
  \label{EQc1Profile}
\end{align}
where $c_1^\rmr(\rv)$ is the one-body direct correlation function that
forms the reference. One exploits that all quantities on the right
hand side of \eqr{EQc1Profile} are either known [$\beta,\mu,
  V_\rmext(\rv)$] or are directly accessible in simulations
[$\rho(\rv)$; potentially via force
  sampling~\cite{rotenberg2020}]. Hence one can construct data for
$c_1^\rmr(\rv)$ to act as a reference.  The one-body direct
correlation function is then transcended to a density functional,
recall: $c_1(\rv;[\rho])=-\delta \beta
F_\rmexc[\rho]/\delta\rho(\rv)$.  The direct correlation learning
\cite{sammueller2023neural, sammueller2023whyNeural} represents
$c_1(\rv;[\rho])$ at a considered position $\rv$ directly as a neural
network, with a simple yet very general multi-layer perceptron
architecture. Supervised training is used to approach
\begin{align}
  c_1(\rv;[\rho]) &= c_1^\rmr(\rv;[\rho]),
  \label{EQc1matching}
\end{align}
where the left hand side is the output of the neural functional and
the right hand side is the reference obtained via \eqr{EQc1Profile}
with simulation input for $\rho(\rv)$. Training to optimize the
matching condition \eqref{EQc1matching} is performed across a range of
(several hundred) training systems and hence differing shapes of the
density profile $\rho(\rv)$. The resulting functional can then be
applied independently of the system size of the original simulations,
thus enabling to make predictions ``beyond-the-box''
\cite{sammueller2023neural, sammueller2023whyNeural}.

\begin{figure*}[!t]
  \vspace{1mm}
  \includegraphics[page=1,width=.99\textwidth]{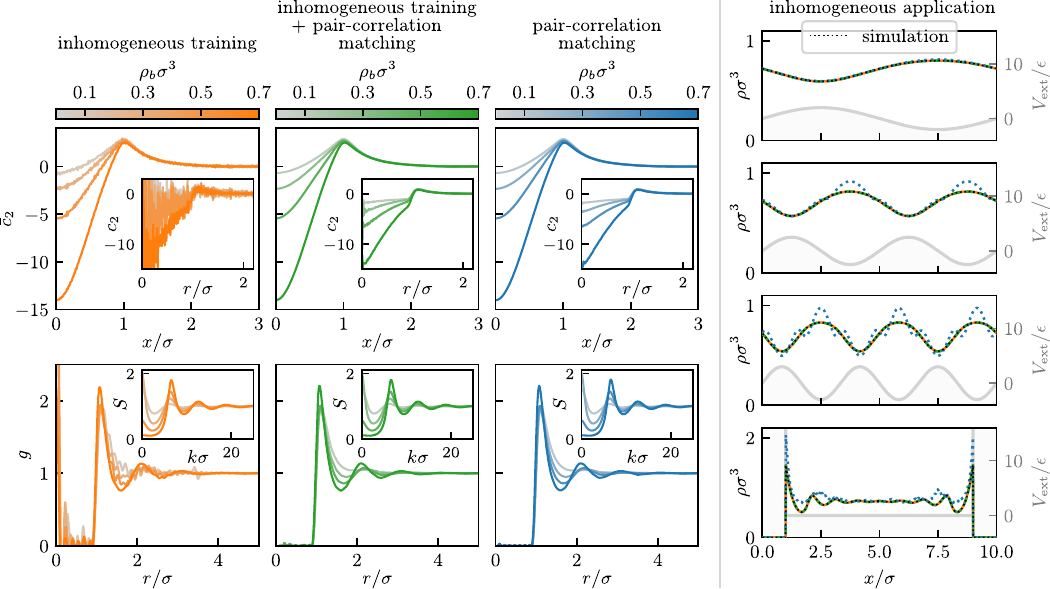}
  \caption{Neural functional results for bulk pair structure and
    planar inhomogeneous density profiles. We show bulk results from
    neural functionals obtained via one-body inhomogeneous training
    (first column), adding the bulk pair correlation regularization
    (second column), and from pure pair-correlation matching (third
    column). The supercritical Lennard-Jones fluid is at reduced
    temperature $k_BT/\epsilon=1.5$ and different scaled bulk
    densities $\rho_b \sigma^3$ (color bar). Shown is the bulk pair
    direct correlation function $\bar{c}_2(x;\rho_b)$ in planar
    geometry as a function of scaled planar distance $x/\sigma$
    obtained via autodiff \eqref{EQc2fromc1} (top panels) and
    $c_2(r;\rho_b)$ as a function of radial distance $r/\sigma$
    (insets in top panels).  The Ornstein-Zernike relation yields the
    corresponding pair correlation function~$g(r)$ (bottom panels) and
    bulk structure factor $S(k)$ (insets in bottom panels).  Also
    shown are predictions from the same three neural functionals for
    planar inhomogeneities (fourth column) as induced by an external
    potential $V_\rmext(x)$ (gray lines) at $\mu/\epsilon=0$. Results
    from the pure pair-correlation matched functional deviate
    increasingly from those with inhomogeneous training, which remain
    almost identical to the simulation reference (thin dotted lines)
    upon increasing inhomogeneity (from top to bottom) even for
    confinement between parallel hard walls (bottom panel).
    \label{FIG1} }
\end{figure*}

Using autodiff to functionally differentiate the one-body direct
correlation functional $c_1(\rv;[\rho])$ yields the two-body direct
correlation functional
\begin{align}
  c_2(\rv,\rv';[\rho]) &= 
  \frac{\delta c_1(\rv;[\rho])}{\delta\rho(\rv')},
  \label{EQc2fromc1}
\end{align}
which is now expressed as a first derivative rather than (and formally
consistent with) the second functional derivative in
\eqr{EQc2fromFexc}.  This implies a significant reduction in terms of
computational complexity for the case of short-ranged (truncated)
interparticle interactions.  Performing the functional
reduction~\eqref{EQc2reduction} then gives results for $c_2(r;\rho_b)$
with genuine predictive status as no information about $c_2^\rmr(r;
\rho_b)$ has entered the local learning scheme. Hence the bulk direct
correlation matching \eqref{EQc2matching} can be used as an {\it a
  posteriori} quality check of the neural functional
$c_1(\rv;[\rho])$.  Similarly, the exchange symmetry
$c_2(\rv,\rv';[\rho])=c_2(\rv', \rv;[\rho])$, which follows from
interchanging the order of derivatives in \eqr{EQc2fromFexc}, is a
non-trivial consistency test.

We have kept the presentation general, but inline with
Refs.~\cite{dijkman2024ml, delasheras2023perspective,
  sammueller2023neural, sammueller2023whyNeural} we train all neural
functionals in planar geometry. We use the supercritical Lennard-Jones
fluid at reduced temperature $k_BT/\epsilon=1.5$ as a generic model
system. Planar geometry offers computational benefits and it
constitutes arguably the most important type of inhomogeneity due to a
multitude of relevant interfacial and substrate applications
(including phase transitions). The technical details of the conversion
from radial to planar geometry are described in our
SI~\cite{sammueller2024pairmatchingSI} and all data is openly
available \cite{sammueller2024pairmatchingGithub}.

To combine the virtues of the different approaches, we first
incorporate the pair-correlation matching \eqref{EQc2matching} into
the local learning method by using \eqr{EQc2fromc1} to express
$c_2(\rv,\rv';[\rho])$ and then proceed to the functional
reduction~\eqref{EQc2reduction}. The loss that is optimized in the
supervised machine learning remains based on the one-body direct
correlation matching \eqref{EQc1matching} but it is supplemented by a
regularizer based on \eqr{EQc2matching}. We find that using the
regularizer improves the training dynamics and the overall quality of
the results achieved, as is desired and expected when supplying
additional information about bulk correlations; details are given in
the SI~\cite{sammueller2024pairmatchingSI}. Note the inhomogeneous
training data is re-used from Ref.~\cite{sammueller2023neural}.  As a
test, we display results for the predicted bulk structure from both
un-regularized one-body learning and pair matching-regularized
one-body learning in Fig.~\ref{FIG1}. The results with regularization
are smoother and the predicted radial distribution functions $g(r)$
are closer to zero inside the core, consistent with the expectation
based on the supply of bulk information into the scheme.

\begin{figure*}[!t]
  \includegraphics[page=1,width=.9\textwidth]{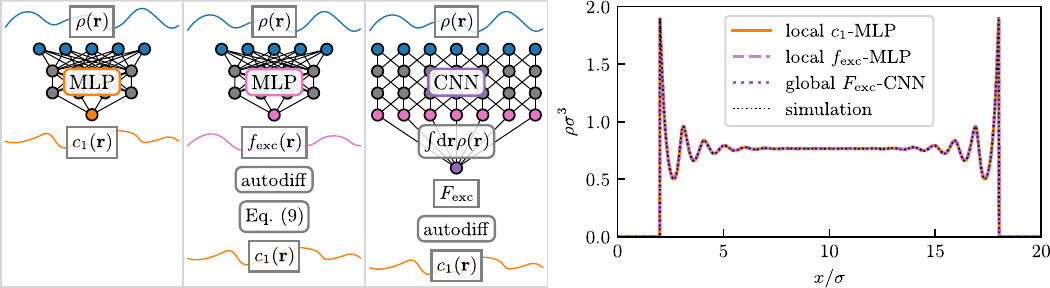}
  \caption{Illustration of the three different network architectures
    for neural functionals. Shown are the multilayer perceptrons (MLP)
    for local learning of $c_1(\rv;[\rho])$ (first column) and for
    $f_\rmexc(\rv;[\rho])$ (second column), and the convolutional
    neural network (CNN) for $F_\rmexc[\rho]$. Corresponding
    predictions for the scaled density profiles $\rho(x)\sigma^3$
    (fourth column) are for the supercritical Lennard-Jones fluid with
    $\mu/\epsilon=1$ and $k_BT/\epsilon=1.5$ confined between two hard
    walls at $x/\sigma=2$ and $x/\sigma=18$. Results for $\rho(x)$ are
    shown from density functional minimization using $c_1(x;[\rho])$
    obtained from local learning according to \eqr{EQc1matching}, from
    local learning of $f_\rmexc(x;[\rho])$ via
    \eqr{EQc1fromLittlefexc}, and from inhomogeneous one-body training
    of the convolutional neural network representation of
    $F_\rmexc[\rho]$. The predictions from the three different neural
    functionals are identical to the simulation reference (thin dotted
    line) on the scale of the plot.
    \label{FIG2} }
\end{figure*}

We have also trained a one-body correlation functional based on
pair-correlation matching alone with no information about
inhomogeneous systems entering. This is potentially important as it
would remove the need to generate inhomogeneous training data, as is
argued in Ref.~\cite{dijkman2024ml}. Within our scheme the method is
straightforward to implement by basing the loss solely on the bulk
direct correlation matching \eqref{EQc2matching}.  Similar to
Ref.~\cite{dijkman2024ml} we provide the required integration
constants by noting from \eqr{EQc1Profile} that $\beta\mu_\rmexc
\equiv \beta \mu - \ln\rho_b = -c_1(\rv;[\rho])$, which the neural
functional delivers with constant density input,
$\rho(\rv)=\rho_b$. Results for the excess chemical potential
$\mu_\rmexc$ are obtained from simulating bulk systems with
$V_\rmext(\rv)=0$. We find despite apparently successful performance
in training metrics that the pair-correlation matched neural
functional $c_1(\rv;[\rho])$ is of significantly poorer quality than
those from the two local one-body learning methods, which rely on
inhomogeneous training data. This becomes clear in predictions for
inhomogeneous systems, where the results of the former neural
functional fall short of those from the inhomogeneously trained
networks, as shown in Fig.~\ref{FIG1}, Col.~4. Also the exchange
symmetry $c_2(\rv,\rv';[\rho])=c_2(\rv',\rv;[\rho])$, which is a
necessary Cauchy-Riemann-like integrability condition for the
existence of a generating functional $F_\rmexc[\rho]$ according to
\eqr{EQc2fromFexc}, is violated, as shown in the SI
\cite{sammueller2024pairmatchingSI}. The method of
Ref.~\cite{dijkman2024ml} satisfies the exchange symmetry by
construction via the neural representation of $F_\rmexc[\rho]$ and the
interchangeability of the order of the two derivatives in
\eqr{EQc2fromFexc}.

We next describe two variants of inhomogeneous one-body learning for
modelling directly the generating functional~$F_\rmexc[\rho]$. We work
with a localized excess free energy integrand $f_\rmexc(\rv;[\rho])$
such that
\begin{align}
  F_\rmexc[\rho] &= \int d\rv \rho(\rv) f_\rmexc(\rv;[\rho]),
  \label{EQFexcAsSpatialIntegral}
\end{align}
where the spatial integration domain is the entire system volume.
The general functional integral $\beta F_\rmexc[\rho]= -\int {\cal
  D}[\rho] c_1(\rv;[\rho])$ is parameterized by
\eqr{EQFexcAsSpatialIntegral} when choosing \cite{evans1992,
  sammueller2023whyNeural}
\begin{align}
   -\beta f_\rmexc(\rv;[\rho]) &=  \int_0^1 da c_1(\rv;[a\rho]).
    \label{EQlocalFreeEnergyAsFunctionalIntegral}
\end{align}
where $a\rho(\rv)$ is a scaled version of the density profile.  The
integrals in Eqs.~\eqref{EQFexcAsSpatialIntegral}
and~\eqref{EQlocalFreeEnergyAsFunctionalIntegral} can be performed
numerically with high efficiency to obtain the value of the excess
free energy~\cite{sammueller2023neural, sammueller2023whyNeural}.

We obtain the corresponding one-body direct correlation functional by
recalling $c_1(\rv;[\rho])=-\delta \beta
F_\rmexc[\rho]/\delta\rho(\rv)$ and inserting $F_\rmexc[\rho]$ in the
form~\eqref{EQFexcAsSpatialIntegral}, which gives
\begin{align}
  c_1(\rv;[\rho]) &= - \beta f_\rmexc(\rv;[\rho]) -
    \int d\rv' \rho(\rv') 
    \frac{\delta  \beta f_\rmexc(\rv;[\rho])}{\delta\rho(\rv')}.
    \label{EQc1fromLittlefexc}
\end{align}
We have exchanged $\rv$ and $\rv'$ according to the symmetry $\delta
f_\rmexc(\rv';[\rho])/\delta\rho(\rv) = \delta
f_\rmexc(\rv;[\rho])/\delta\rho(\rv')$; this identity follows from
differentiating \eqr{EQlocalFreeEnergyAsFunctionalIntegral} and from
the exchange symmetry of $c_2(\rv,\rv';[\rho])$.  The form
\eqref{EQc1fromLittlefexc} is suitable for the application to local
learning.

In our first free energy method we represent $f_\rmexc(\rv;[\rho])$ as
a neural functional and perform one-body direct correlation matching
via~\eqr{EQc1matching} on the basis of \eqr{EQc1fromLittlefexc}, where
autodiff is used to carry out the functional derivative. The second
free energy method uses a convolutional neural network architecture
similar to that of Ref.~\cite{dijkman2024ml} to implement via
$f_\rmexc(\rv;[\rho])$ the excess free energy functional
$F_\rmexc[\rho]$ in the structure of \eqr{EQFexcAsSpatialIntegral}.
We thereby perform no pooling (coarse-graining, as is common practice
\cite{chollet2017}) and design a final layer that represents
$f_\rmexc(\rv;[\rho])$ which is then multiplied by~$\rho(\rv)$ and
integrated according to \eqr{EQFexcAsSpatialIntegral} for the global
value of $F_\rmexc[\rho]$; details are given in the
SI~\cite{sammueller2024pairmatchingSI}.  This specific convolutional
neural network design extends naturally to arbitrary system sizes,
thus replicating the capability of the local correlation learning for
making predictions ``beyond-the-box'' \cite{sammueller2023neural,
  sammueller2023whyNeural}.  Illustrations of the different neural
functionals together with representative results from the three
methods for planar density profiles~$\rho(x)$ of a Lennard-Jones fluid
confined between parallel hard walls are shown Fig.~\ref{FIG2}. Having
been trained with the same inhomogeneous one-body profiles from grand
canonical Monte Carlo simulations \cite{sammueller2023neural}, all
methods give an excellent account of the highly inhomogeneous spatial
fluid structure.

We conclude with several conceptual points related to the inherent
functional reduction \eqref{EQc2reduction} of bulk pair-correlation
matching \cite{dijkman2024ml}. The feeding of knowledge of~$g(r)$ into
the neural functional constitutes a very significant amount of
information about the system under consideration. Henderson's
uniqueness theorem \cite{henderson1974uniqueness} formally ensures
that for systems interacting solely with a pair potential $\phi(r)$,
perfect knowledge of $g(r)$, at a single statepoint, is sufficient to
determine $\phi(r)$.  Density functional theory \cite{evans1979} then
implies that $F_\rmexc[\rho]$ is uniquely determined in principle,
given the information that no higher-body interparticle interactions
are present \cite{evans1990reverseMC}.

However, it is difficult to see why a neural functional trained
through the aperture of the functional reduction~\eqref{EQc2reduction}
would render Henderson's theorem operational. As the density
functional framework \eqref{EQc2fromFexc}--\eqref{EQc2matching},
utilized within the pair-correlation matching, applies equally to a
system with many-body interparticle interactions, we see no mechanism
in Ref.~\cite{dijkman2024ml} that would intrinsically reduce
$F_\rmexc[\rho]$ to pair interactions only (see
Ref.~\cite{evans1990reverseMC} for a related discussion in the context
of simulation work). This situation cannot be remedied by supplying
$g(r)$ in more abundance and with higher precision.  In contrast the
one-body matching condition \eqref{EQc1matching} allows for full
exploration of the entire density functional dependence for any given
form of the underlying Hamiltonian, which gives a formal mechanism to
rationalize the high quality of our correspondingly obtained results.
Our locally trained functional $c_1(\rv;[\rho])$
\cite{sammueller2024pairmatchingGithub} does not suffer from the
somewhat limited extrapolation capabilities reported for the
inhomogeneous training method of Ref.~\cite{dijkman2024ml}.
We also find robust values when evaluating $c_1(\rv;[\rho])$ at
positions~$\rv$ inside of a hard wall, where the local density
vanishes, which corroborates the extrapolation capability to unseen
cases.  Such regions do per construction not contribute directly to
the free energy integral \eqref{EQFexcAsSpatialIntegral}, due to the
multiplication with $\rho(\rv)=0$.

In future work it would be interesting to further compare the
  different one-body learning strategies to each other.
One could also attempt to utilize test particle concepts
\cite{percus1962} to identify $\rho(\rv) = \rho_b g(r)$ upon setting
$V_\rmext(\rv)=\phi(r)$, as also explored dynamically
\cite{treffenstaedt2021dtpl, treffenstaedt2022dtpl, schmidt2022rmp}
and quantum mechanically \cite{mccary2020prlBlueElectron}.
Despite the implied restriction to spherical symmetry, this would
allow simulation results for pair correlation functions $g(r)$ to be
used as genuinely inhomogeneous one-body density profiles
$\rho(\rv)$.  The three-body direct correlation functional
follows from (Hessian) autodiff with respect to the density
profile according to $c_3(\rv,\rv',\rv'';[\rho]) = \delta
c_1(\rv;[\rho])/[\delta\rho(\rv')\delta\rho(\rv'')]$
\cite{sammueller2023neural}. Then one could exploit Noether
invariance sum rules that relate to spatial gradients of
$c_2(\rv,\rv';[\rho])$ \cite{sammueller2023whyNeural,
hermann2021noether}. Furthermore the use of the force density
integral $\rho(\rv)\nabla c_1(\rv;[\rho])=-\int d\rv'
\rho_2(\rv,\rv')\nabla\beta\phi(|\rv-\rv'|)$ \cite{hansen2013,
  evans1979, yvon1935, born1946} could be beneficial for making
progress despite an increase in complexity \cite{tschopp2022forceDFT,
  sammueller2022forceDFT}.  Localized forces are also relevant for
functional treatments of general observables
\cite{sammueller2024hyperDFT} and for nonequilibrium flow
\cite{delasheras2023perspective, zimmermann2024ml}.

\smallskip

\begin{acknowledgments}
  We thank Jacobus Dijkman, Max Welling, Jan-Willem van de Meent,
  Bernd Ensing, Kieron Burke, Marjolein Dijkstra, Bob Evans, and
  Stefanie Kampa for useful discussions.
\end{acknowledgments}

%% \bibliographystyle{prsty}
%% \bibliography{noe}



\section*{Supplementary material}

# Supplementary Information for
# "Neural density functionals: Local learning and pair-correlation matching"

Florian Sammüller and Matthias Schmidt

*Theoretische Physik II, Physikalisches Institut, Universität Bayreuth, D-95447 Bayreuth, Germany*[*]

(Dated: July 25, 2024)

## I. LOCAL INHOMOGENEOUS ONE-BODY LEARNING

### A. Training data

Training of the neural correlation functional proceeds as in Ref. [1] with GCMC simulation data for inhomogeneous randomized external potentials. We consider the truncated ($r_c = 2.5\sigma$) Lennard-Jones fluid at constant supercritical [2] temperature $k_B T/\epsilon = 1.5$. We reuse the training data set of Ref. [1] consisting of 500 simulation results in a box of length $20\sigma$ in the inhomogeneous $x$-direction and lateral lengths of $10\sigma$. The inhomogeneities in the considered training set are substantial and result in local density maxima of up to $\rho(x)\sigma^3 \lesssim 6$, which nevertheless poses no difficulty during sampling (in contrast to accessing the radial distribution function $g(r)$ for large bulk densities, cf. Sec. III A). The data set is deposited in Zenodo [3].

One-body direct correlation profiles are calculated pointwise from the sampled density profiles $\rho(x)$ according to

$$c_1^{\text{ref}}(x;[\rho]) = \ln \rho(x) + \beta V_{\text{ext}}(x) - \beta\mu, \qquad (1)$$

where the external potential $V_{\text{ext}}(x)$, chemical potential $\mu$ and inverse temperature $\beta = 1/(k_B T)$ are known input quantities of the simulations. The logarithm in Eq. (1) is understood as $\ln(\rho(x)\Lambda^3)$, where the thermal wavelength $\Lambda$ is set to the particle size $\sigma$, which defines our unit of length.

### B. Neural network and training procedure

The neural network used for the local representation of the one-body direct correlation functional $c_1(x;[\rho])$ possesses a straightforward multilayer perceptron (MLP) architecture, see Fig. S5. We use three hidden layers with 512 nodes and smooth non-linear activation functions (Gaussian error linear units, "GELU"). The input layer consists of 701 nodes, as the density profile is provided in a window of size $3.5\sigma$ around the location of interest with a discretization interval of $\Delta x = 0.01\sigma$.

---

[*] Florian.Sammueller@uni-bayreuth.de

Training the neural network on the basis of inhomogeneous one-body learning amounts to minimizing the loss

$$L_{\text{inhom}} = \sum_{i,j} \left(c_1(x_i;[\rho_j]) - c_1^{\text{ref}}(x_i;[\rho_j])\right)^2, \qquad (2)$$

where the neural network prediction $c_1(x_i;[\rho_j])$ is compared to the reference data $c_1^{\text{ref}}(x_i;[\rho_j])$ for each discretized location $x_i$ of every inhomogeneous system $j$ in the training set. Code, simulation data and trained models are openly available [4].

### C. Applications

For the prediction of density profiles, rearranging Eq. (1) yields

$$\rho(x) = \exp\left(-\beta(V_{\text{ext}}(x) - \mu) + c_1(x;[\rho])\right), \qquad (3)$$

which constitutes the standard Euler-Lagrange equation of DFT. Given an analytical or neural representation of $c_1(x;[\rho])$, Eq. (3) can be solved self-consistently for $\rho(x)$ with a standard Picard iteration [1].

Accessing structural information on the pair-correlation level is facilitated by functional differentiation of $c_1(x;[\rho])$, which yields the two-body direct correlation functional

$$c_2(x, x';[\rho]) = \frac{\delta c_1(x;[\rho])}{\delta \rho(x')}. \qquad (4)$$

The functional derivative can be evaluated efficiently for general density input $\rho(x)$ by automatic differentiation and normalization with the discretization interval $\Delta x$. Using automatic differentiation is particularly suitable and natural in case of a neural-network-based representation of $c_1(x;[\rho])$.

## II. BULK PAIR-CORRELATIONS

The considered neural functionals operate on the level of direct correlations in planar geometry. In the following, we lay out conversions between radial and planar geometry in direct and Fourier space. The Ornstein-Zernike equation is utilized to obtain total correlation functions from neural predictions and, inversely, to convert radial distribution functions $g(r)$ from direct measurement in simulation to corresponding direct correlation functions for use in the pair-correlation matching, see Sec. III A.

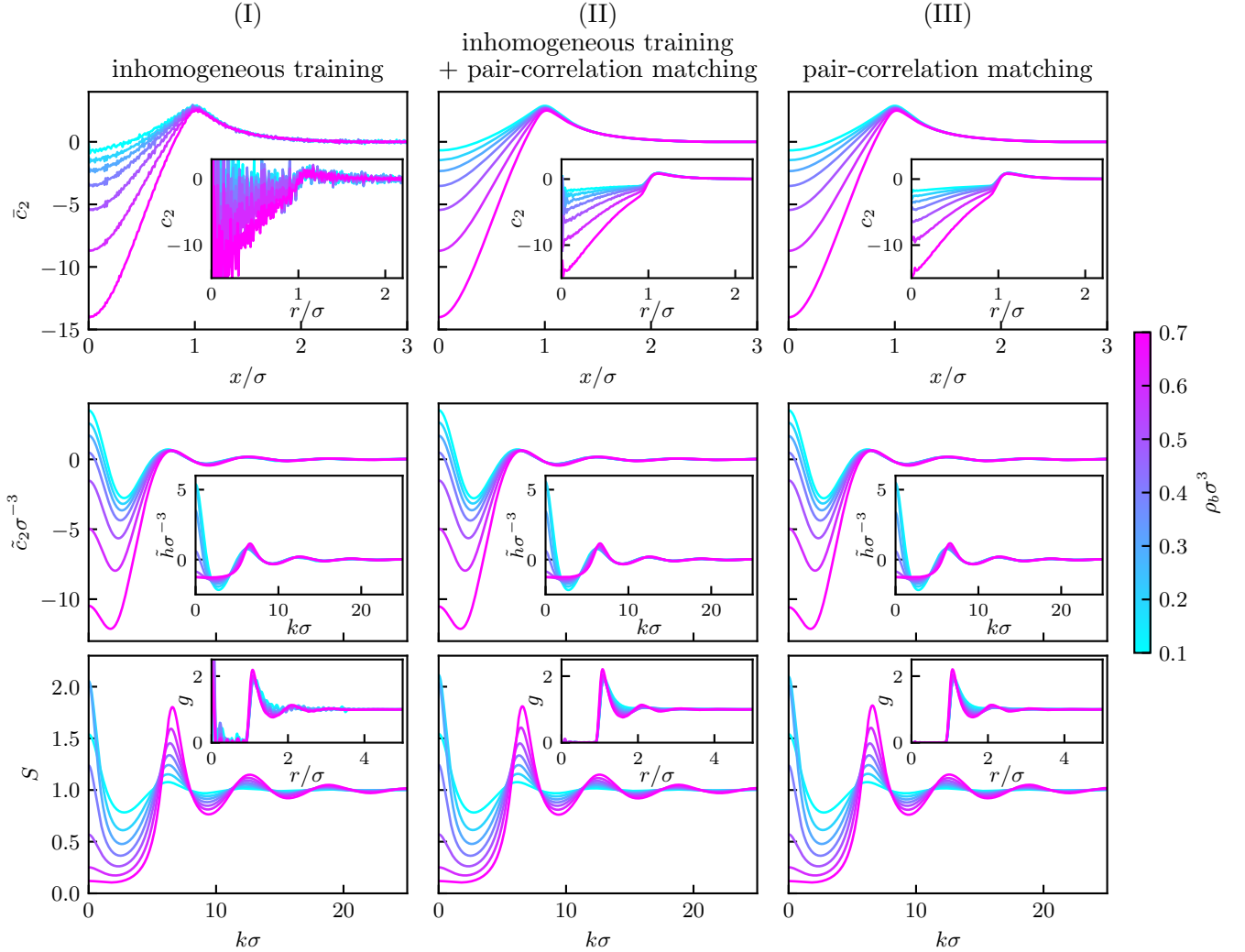

FIG. S1. Comparison of predictions of neural functionals for the truncated Lennard-Jones fluid at temperature $k_B T/\epsilon = 1.5$ and bulk densities $\rho_b \sigma^3 = 0.1, 0.2, 0.3, 0.4, 0.5, 0.6, 0.7$. Results are shown for training with purely inhomogeneous reference data (I, left column), with additional pair-correlation regularization (II, middle column) as well as with pure pair-correlation matching (III, right column). We depict the planar two-body direct correlation function $\bar{c}_2(x; \rho_b)$ and its radial representation $c_2(r; \rho_b)$ (first row), the direct and total correlation functions $\tilde{c}_2(k; \rho_b)$ and $\tilde{h}(k; \rho_b)$ in Fourier space (second row), as well as the static structure factor $S(k; \rho_b)$ and the radial distribution function $g(r; \rho_b)$ (third row).

## A. Planar and radial geometry

The planar projection of a generic radially symmetric function $f(\mathbf{r}) = f(|\mathbf{r}|) = f(r)$ is given by

$$\bar{f}(x) = \int dy \int dz \, f(\sqrt{x^2 + y^2 + z^2}) = 2\pi \int_x^\infty dr \, r f(r), \quad (5)$$

where we indicate quantities in the reduced planar geometry by an overbar. By differentiating Eq. (5), one arrives at the inverse transformation from planar to radial geometry via

$$f(r) = -\frac{1}{2\pi r} \left.\frac{\partial \bar{f}(x)}{\partial x}\right|_{x=r}, \quad (6)$$

where we have assumed radial symmetry for $f(r)$. Applying a one-dimensional Fourier transform to the planar projection of a radially symmetric function yields directly the radially symmetric representation in Fourier space (indicated by a tilde),

$$\int dx \, \bar{f}(x) \exp(i k_x x) = \int d\mathbf{r} \, f(r) \exp(i\mathbf{k} \cdot \mathbf{r}) \\ = \tilde{f}(|\mathbf{k}|) = \tilde{f}(k), \quad (7)$$

where $\int d\mathbf{r}$ denotes a volume integral over the entire domain.

Transforming a radially symmetric function $f(r)$ in real space to its radially symmetric Fourier representation $\tilde{f}(k)$ is facilitated by a radial Fourier(-Hankel) trans-

form,

$$\tilde{f}(k) = \frac{4\pi}{k} \int_0^\infty dr\, r f(r) \sin(kr). \quad (8)$$

The inverse transform of Eq. (8) is identical up to a prefactor,

$$f(r) = \frac{1}{2\pi^2 r} \int_0^\infty dk\, k \tilde{f}(k) \sin(kr). \quad (9)$$

The numerical evaluation of Eqs. (8) and (9) can be facilitated either via a fast Fourier transform (FFT) or via a discrete sine transform (DST). In both cases, particular care must be taken to account correctly for the discretization and the implied periodicity of the radially symmetric function to be transformed.

### B. Direct and total correlation functions

For the considered two-body direct correlation function $\bar{c}_2(x; \rho_b)$, as obtained from the neural functional by autodifferentiation (4) and network evaluation in bulk with constant density input $\rho(x) = \rho_b$, Eq. (6) gives access to the more common radial representation $c_2(r; \rho_b)$, as shown in Figs. 1 and S1. Due to the required numerical differentiation and the division by $r$ in Eq. (6), the results for $c_2(r; \rho_b)$ are prone to numerical artifacts in particular for small values of $r$.

Commencing with a neural functional prediction for $\bar{c}_2(x; \rho_b)$, a one-dimensional Fourier transform yields the one-body direct correlation function $\tilde{c}_2(k; \rho_b)$ in radial geometry according to Eq. (7). The total correlation function $\tilde{h}(k; \rho_b)$ can then be calculated algebraically in Fourier space via the Ornstein-Zernike equation

$$\tilde{h}(k; \rho_b) = \frac{\tilde{c}_2(k; \rho_b)}{1 - \rho_b \tilde{c}_2(k; \rho_b)}. \quad (10)$$

From $\tilde{h}(k; \rho_b)$, one obtains the static structure factor

$$S(k; \rho_b) = 1 + \rho_b \tilde{h}(k; \rho_b), \quad (11)$$

and the transformation (9) back to direct space gives access to the radial distribution function

$$g(r; \rho_b) = h(r; \rho_b) + 1. \quad (12)$$

Neural functional predictions of the quantities in Eqs. (10)–(12) are shown in Fig. S1 at different bulk densities $\rho_b \sigma^3 = 0.1, 0.2, 0.3, 0.4, 0.5, 0.6$, and $0.7$.

### III. PAIR-CORRELATION MATCHING AND REGULARIZATION

### A. Preparation of simulation data

We have performed 500 simulations of the Lennard-Jones fluid in a cubic box of size $(20\sigma)^3$ at different chemical potentials chosen uniformly in the range $-7 < \mu/\epsilon < 0$. This yields accurately sampled radial distribution functions $g(r; \rho_b)$ in $0 \leq r/\sigma \leq 10$ for bulk densities $0.01 \lesssim \rho_b \sigma^3 \lesssim 0.72$. Accessing larger bulk densities is hindered by the fact that $g(r)$ still displays non-negligible oscillations for $\rho_b \sigma^3 \gtrsim 0.72$ for the maximal radial distance $r = 10\sigma$ in the considered simulation box. This stands in contrast to the sampling of inhomogeneous density profiles, where much larger local density maxima pose no substantial difficulty, see Sec. I A.

From the simulation data for $g(r; \rho_b)$, we calculate:

- $h(r; \rho_b)$ via the definition (12) of the total pair-correlation function,
- $\tilde{h}(k; \rho_b)$ via the radial Fourier transform (8),
- $\tilde{c}_2(k; \rho_b)$ via the Ornstein-Zernike equation (10),
- $c_2(r; \rho_b)$ via the inverse radial Fourier transform (9),
- $\bar{c}_2(x; \rho_b)$ via the planar projection (5).

The planar bulk two-body direct correlation function $\bar{c}_2(x; \rho_b)$ can be compared directly to the functional derivative (4) of the one-body direct correlation functional in planar geometry,

$$\bar{c}_2(x; \rho_b) = \left.\frac{\delta c_1(0; [\rho])}{\delta \rho(x)}\right|_{\rho=\rho_b}. \quad (13)$$

Automatic differentiation is used to evaluate the functional derivative of the neural-network-based representation of $c_1(x; [\rho])$ in Eq. (13).

### B. Loss functions and training procedure

Pair-correlation matching is implemented via the loss

$$L_{\rm pc} = \sum_{i,j} \left(\bar{c}_2(x_i; \rho_{b,j}) - \bar{c}_2^{\rm ref}(x_i; \rho_{b,j})\right)^2 \\ + \sum_j \left(\mu_{\rm exc}(\rho_{b,j}) - \mu_{\rm exc}^{\rm ref}(\rho_{b,j})\right)^2, \quad (14)$$

where $\bar{c}_2(x_i; \rho_{b,j})$ is calculated for the given bulk density input $\rho_{b,j}$ of bulk simulation $j$ according to Eq. (13). The result is compared directly to the pre-calculated simulation reference $\bar{c}_2^{\rm ref}(x_i; \rho_{b,j})$, see Sec. III A, for each discretized spatial coordinate $x_i$. To fix the remaining integration constant, the excess chemical potential $\mu_{\rm exc}(\rho_b) = -k_B T c_1(x; \rho_b)$ of the neural network prediction is compared to the simulation reference $\mu_{\rm exc}^{\rm ref}$, which serves as an additive contribution to $L_{\rm pc}$.

In general, we consider a linear combination of both loss functions,

$$L = \alpha_{\rm inhom} L_{\rm inhom} + \alpha_{\rm pc} L_{\rm pc}, \quad (15)$$

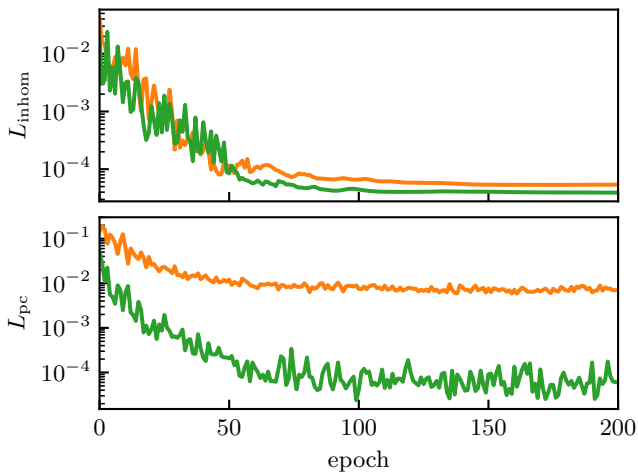

FIG. S2. Training dynamics with pair-correlation regularization (II, green) results in lower loss values $L_{\text{inhom}}$ and $L_{\text{pc}}$ as compared to pure inhomogeneous one-body learning without regularization (I, orange). We show the $L_{\text{inhom}}$ and $L_{\text{pc}}$ as a function of the training epoch.

which is used for the backpropagation and for the adaptation of the trainable parameters of the neural network. The constant factors $\alpha_{\text{inhom}}$ and $\alpha_{\text{pc}}$ control the relative influence of inhomogeneous one-body matching (2) and pair-correlation matching (14) by weighting the respective loss terms. Three choices are considered:

(I) $\alpha_{\text{inhom}} = 1$, $\alpha_{\text{pc}} = 0$:
purely inhomogeneous one-body learning as in Ref. [1],

(II) $\alpha_{\text{inhom}} = 1$, $\alpha_{\text{pc}} = 0.01$:
inhomogeneous one-body learning with pair-correlation regularization as introduced in this work,

(III) $\alpha_{\text{inhom}} = 0$, $\alpha_{\text{pc}} = 1$:
pure pair-correlation matching inspired by Ref. [5].

### C. Training dynamics and bulk results

Fig. S2 shows a comparison of the training dynamics for the individual loss terms $L_{\text{inhom}}$ and $L_{\text{pc}}$ for inhomogeneous one-body learning with (II) and without (I) pair-correlation regularization. The magnitude of both loss terms is decreased for the case of added pair-correlation regularization.

In Fig. S1, we depict results for bulk pair-correlation functions for each training strategy (I)–(III). Purely inhomogeneous one-body training (I) yields slightly noisy derivatives $\bar{c}_2(x;\rho_b)$, which hamper the performance of challenging numerical transformations, e.g. to obtain the two-body direct correlation function $c_2(r;\rho_b)$ in radial geometry or the radial distribution function $g(r)$. This problem is resolved both with the added pair-correlation

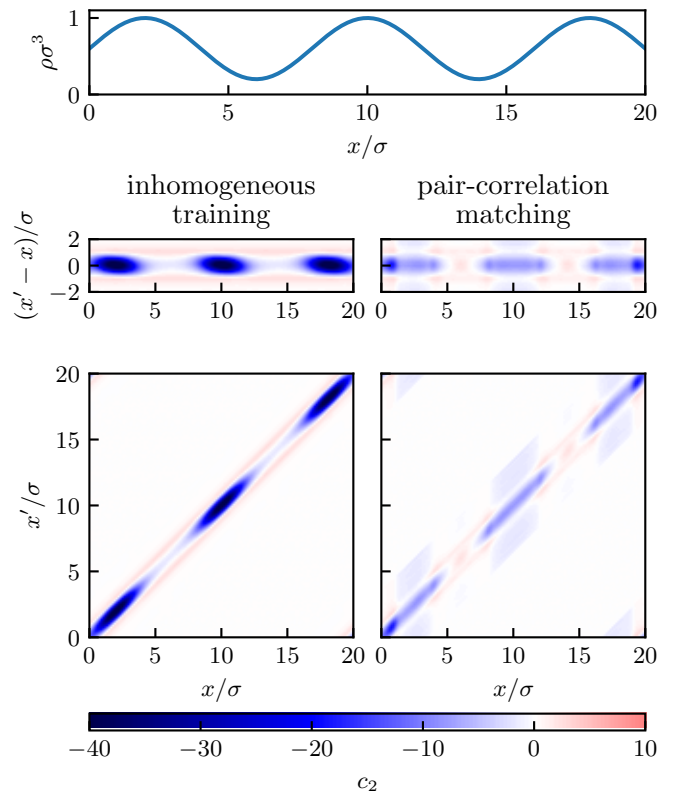

FIG. S3. Two-body direct correlation functional $c_2(x, x'; [\rho])$ for an exemplary inhomogeneous density profile $\rho(x)$ as input. We show results for neural functionals that have been trained via inhomogeneous one-body learning (I) and with pure pair-correlation matching (III). The results of the pair-correlation-matched neural functional violate the expected exchange symmetry $c_2(x, x'; [\rho]) = c_2(x', x; [\rho])$.

regularization (II) as well as with pure pair-correlation matching (III). Although the latter strategy enables accurate bulk predictions due to the supply of bulk data during training, it provides no reasonable extrapolation to inhomogeneous scenarios, as seen e.g. in inaccurate predictions of density profiles (Fig. 1) and of inhomogeneous two-body correlations (Fig. S3).

## IV. INHOMOGENEOUS TWO-BODY CORRELATIONS AND NOETHER SUM RULES

Autodifferentiation also gives access to $c_2(x, x'; [\rho])$ for inhomogeneous density input according to Eq. (4), which enables further quality assessments of the neural functionals. As the two-body direct correlation functional formally arises as a second-order functional derivative of the excess free energy functional $F_{\text{exc}}[\rho]$, testing the expected exchange symmetry $c_2(x, x'; [\rho]) = c_2(x', x; [\rho])$ serves as a first consistency check. Additionally, testing

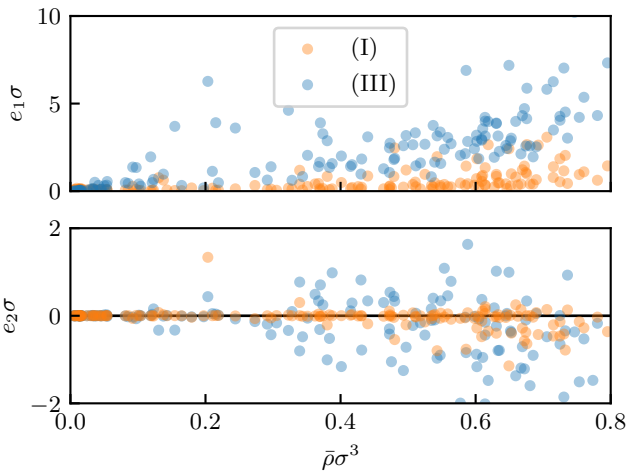

FIG. S4. Test of the Noether sum rules (16) and (17) as obtained from neural functionals trained on the basis of inhomogeneous one-body profiles (I, orange) and via bulk pair-correlation matching (III, blue). The respective errors $e_1$ and $e_2$ of the predictions are considerably larger in the latter case.

the validity of the sum rules

$$\nabla c_1(\mathbf{r};[\rho]) = \int d\mathbf{r}'\, c_2(\mathbf{r},\mathbf{r}';[\rho])\nabla'\rho(\mathbf{r}'), \qquad (16)$$

$$\int d\mathbf{r}\, \rho(\mathbf{r}) \int d\mathbf{r}'\, \rho(\mathbf{r}')\nabla c_2(\mathbf{r},\mathbf{r}';[\rho]) = 0, \qquad (17)$$

which result from thermal Noether invariance [6], constitutes a further valuable test of the fitness of a neural functional [1].

We depict in Fig. S3 results for $c_2(x,x';[\rho])$ with inhomogeneous density input as obtained from neural functionals that have been trained solely with inhomogeneous one-body data (I) and with bulk pair-correlations (III), respectively. The results from the pair-correlation-matched neural functional do not display the expected symmetry of the two-body direct correlation functional upon exchanging the spatial coordinates. We recall that within the local learning strategy, the exchange symmetry is not automatically enforced, unlike in Ref. [5] where the global excess free energy acts as the output of their neural network.

In Fig. S4, we show the discrepancy of the Noether sum rules (16) and (17) as defined via $e_1 = \|\partial_x c_1(x;[\rho]) - \int dx'\, c_2(x,x';[\rho])\partial_{x'}\rho(x')\|_\infty$ and $e_2 = \int dx\, \rho(x) \int dx'\, \rho(x')\partial_x c_2(x,x';[\rho])$. The results of the neural functional from pure pair-correlation matching (III) show considerable deviations from these sum rules.

## V. NEURAL FREE ENERGY METHODS

As an alternative to working with one-body direct correlations, we also consider neural networks that give immediate access to the excess free energy functional $F_{\text{exc}}[\rho]$. We base the construction of these neural free energy functionals on the parametrization

$$F_{\text{exc}}[\rho] = \int d\mathbf{r}\, \rho(\mathbf{r}) f_{\text{exc}}(\mathbf{r};[\rho]), \qquad (18)$$

which provides a well-defined localization of the excess free energy via the uniquely determined quantity

$$f_{\text{exc}}(\mathbf{r};[\rho]) = -k_B T \int_0^1 da\, c_1(\mathbf{r};[a\rho]). \qquad (19)$$

### A. Local learning of $f_{\text{exc}}(\mathbf{r};[\rho])$

As a first approach, it is hence natural to construct a local neural-network-based representation of $f_{\text{exc}}(\mathbf{r};[\rho])$. We specialize to the considered planar geometry and reuse the MLP architecture of the neural correlation functional as described in Sec. I B and shown in Fig. S5.

Training is facilitated by inhomogeneous one-body matching (2), where one determines

$$c_1(\mathbf{r};[\rho]) = -\beta f_{\text{exc}}(\mathbf{r};[\rho]) - \int d\mathbf{r}'\, \rho(\mathbf{r}')\frac{\delta\beta f_{\text{exc}}(\mathbf{r};[\rho])}{\delta\rho(\mathbf{r}')}, \tag{20}$$

with the functional derivative of $f_{\text{exc}}(\mathbf{r};[\rho])$ being obtained by automatic differentiation within the training loop. To be able to evaluate the integral in a local learning scheme, the arguments $\mathbf{r}$ and $\mathbf{r}'$ have been swapped according to the expected symmetry $\delta f_{\text{exc}}(\mathbf{r};[\rho])/\delta\rho(\mathbf{r}') = \delta f_{\text{exc}}(\mathbf{r}';[\rho])/\delta\rho(\mathbf{r})$ that arises from Eq. (19) and $c_2(\mathbf{r},\mathbf{r}';[\rho]) = c_2(\mathbf{r}',\mathbf{r};[\rho])$. Crucially, this symmetry is not enforced intrinsically for an untrained neural functional for $f_{\text{exc}}(\mathbf{r};[\rho])$, but it rather arises only after sufficient training. Utilizing the interchangeability of $\mathbf{r}$ and $\mathbf{r}'$ in Eq. (20) for the calculation of one-body direct correlations during training is nevertheless valid, as this leaves the true minimum of the matching condition (2) unchanged.

Applying the local neural functional for the prediction of free energy values is straightforward via the numerical evaluation of Eq. (18) for the considered density profile $\rho(x)$. For the prediction of equilibrium density profiles, $c_1(x;[\rho])$ is obtained in the considered planar geometry from Eq. (20) for use in the self-consistent solution of Eq. (3).

### B. Convolutional neural network for $F_{\text{exc}}[\rho]$

As an alternative to the local learning of $f_{\text{exc}}(\mathbf{r};[\rho])$, we use a convolutional neural network (CNN) to represent the global quantity $F_{\text{exc}}[\rho]$ while retaining the specific structure of Eq. (18). This approach bears similarity to the neural network construction in Ref. [5], but crucially differs in the utilization of a specific parametrization, Eq. (18), and in the chosen architecture, which is described in the following and illustrated in Fig. S6.

The CNN consists of 8 convolutional layers with kernel size 11, [16, 16, 32, 32, 64, 64, 16, 1] filters and dilation rates of [1, 2, 4, 8, 16, 32, 64, 1]. These hyperparameters are chosen to enable the propagation of non-local density information in a range that is similar to the window widths of the MLPs used in the previous local learning methods. We use cyclic padding for the convolutions, as is appropriate for the periodic boundary conditions of the considered systems, and employ no pooling or other coarse-graining layers, thus keeping full spatial resolution throughout the network. As a consequence of this fully convolutional architecture, the network remains applicable to virtually arbitrary system sizes and hence facilitates predictions "beyond-the-box" [1], as is also the case for the local learning approaches.

To arrive at the value of $F_{\text{exc}}[\rho]$ in the spirit of Eq. (18), the last layer is multiplied pointwise with the input density profile $\rho(\mathbf{r})$ and integrated over the whole domain. In particular, this implies that this last convolutional layer acts as a representation of the one-body profile $f_{\text{exc}}(\mathbf{r}; [\rho])$.

For application in the self-consistent calculation of density profiles, automatic differentiation of the $F_{\text{exc}}[\rho]$-CNN with respect to the input density $\rho(\mathbf{r})$ yields the complete one-body direct correlation profile $c_1(\mathbf{r}; [\rho])$ to be used in Eq. (3).

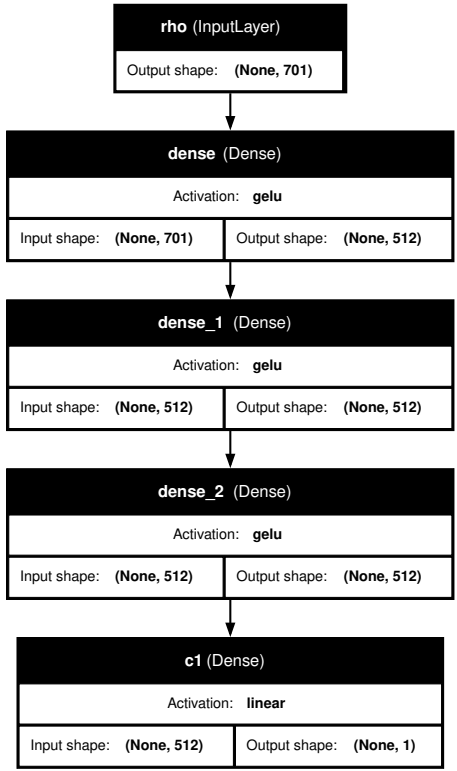

FIG. S5. Architecture of the MLP for local learning of $c_1(x;[\rho])$, cf. Sec. I B, and $f_{\text{exc}}(x;[\rho])$, cf. Sec. V A. The input and output shapes indicate the variable batch size ("None") and the number of nodes for each layer.

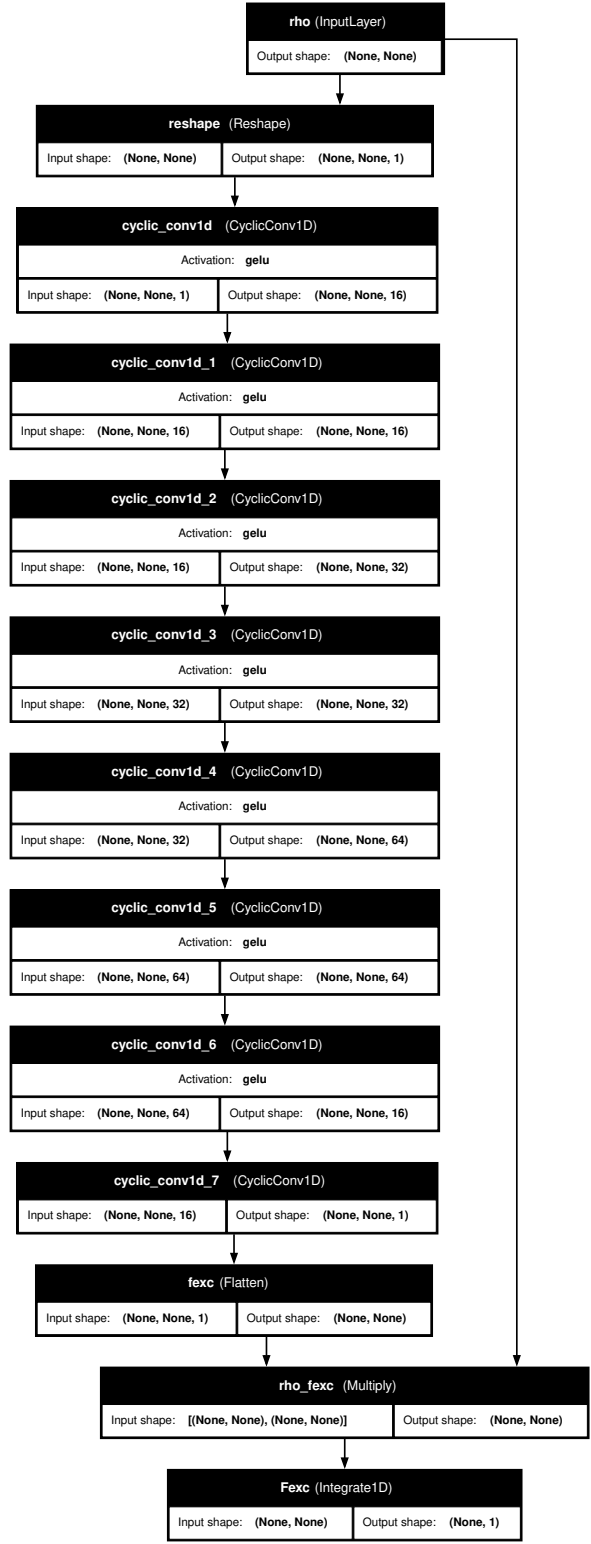

FIG. S6. Architecture of the CNN for $F_{\text{exc}}[\rho]$, cf. Sec. V B. The input and output shapes indicate the batch size, the number of spatial bins ("None" for variable size) and the number of channels.



\end{document}